\documentclass[11pt]{article}

\usepackage[margin=1in]{geometry}
\usepackage{microtype}
\usepackage{graphicx}
\usepackage{booktabs}
\usepackage{longtable}
\usepackage{amsmath, amssymb, mathtools}
\usepackage{enumitem}
\usepackage{hyperref}
\usepackage{caption}
\usepackage{tikz}
\usetikzlibrary{positioning,arrows.meta}

\usepackage[round]{natbib}

\hypersetup{
  colorlinks=true,
  linkcolor=blue,
  citecolor=blue,
  urlcolor=blue
}

\title{Computable Structuralism: A Categorical Rewrite Calculus of Mythic Variants}
\author{Juan J. Segura, Universidad Andres Bello, Santiago, Chile}
\date{}

\begin{document}
\maketitle

\begin{abstract}
Structural approaches to myth and narrative are compelling in close reading but hard to compare across traditions, media, and scale. We propose a formal framework that renders Lévi-Straussian transformation as mathematics while remaining readable as narrative analysis. Variants, superhero continuities, and franchise arcs are modeled as typed rewrite programs on a coupled two-register state $(X,Y)$, abstracting an everyday/social channel and a symbolic/legitimation channel. The canonical formula becomes coherence data: a natural transformation $\eta:U\Rightarrow V$ between update endofunctors, where $U$ updates each register in place and $V$ performs a swap+inversion. Context is internalized by operator choice, turning naturality into a corpus-facing type check: failures diagnose mis-specified oppositions or illegal transport; successes witness coherent structural models. Order effects are summarized by a five-value invariant (Key). We apply the method to 80 narratives (20 folktales, 20 religious myths, 20 superheroes, 20 franchises), each encoded as $(a,b,x,y)$ with a Key. 59/80 (74\%) explicitly name a normative constraint in $y$ (law, taboo, contract, prophecy), supporting the two-register abstraction. The result is a testable bridge between structural anthropology and cultural analytics: stories remain interpretable yet become transportable objects for computation, comparison, and falsifiable constraints on transformation.
\end{abstract}


\section{Introduction}
Comparative analysis of myths, folktales, and popular narrative cycles is often pulled between two methodological impulses. One insists that meaning is local: the social, ritual, and historical conditions of a narrative are not detachable context but part of the object itself. The other insists that comparison is possible because narratives are not isolated texts but \emph{variants related by transformations}—substitutions, inversions, role exchanges, and systematic reassignments of value. Structural anthropology sharpened this second impulse by treating myths less as messages and more as systems of transformation whose constraints can be stated independently of any single telling \citep{LeviStraussStructuralStudyMyth,LeviStraussStructuralAnthropology}.

In contemporary cultural analytics, a third pressure appears \citep{Manovich2007CulturalAnalytics}. If comparison is to scale—even modestly—methods must become explicit enough to reproduce, contest, and improve. This is not a demand for full automation; it is a demand for \emph{inspectable form}. ``Operationalizing Lévi-Strauss'' in this setting means specifying (i) what the comparable objects are, (ii) what counts as a lawful transformation between them, and (iii) what invariants remain stable under those transformations. Without these commitments, structural language risks becoming a rhetoric of inevitability: the analyst can always ``see'' an inversion, but the corpus cannot resist the claim.

This article proposes a framework designed to satisfy contextual sensitivity, transformational comparability, and explicit operational form at once. The core move is to represent narratives as \emph{typed rewrite programs} acting on a coupled two-register configuration $(X,Y)$. The registers implement a pragmatic division that repeatedly surfaces across traditions and media. The $X$ register tracks the everyday/social channel—agents, institutions, survival constraints, political conflict, and material conditions. The $Y$ register tracks the symbolic/legitimation channel—law, taboo, prophecy, covenant, ritual, moral constraint, and cosmological order. The claim is not metaphysical (it is not a theory of mind); it is operational: many narratives treat transformations in $X$ as consequential only when they are recognized, authorized, prohibited, or punished in $Y$, and conversely treat disruptions in $Y$ as real only when they reorganize $X$.

To connect this representation to canonical structural transformation, the paper treats Lévi-Strauss’s canonical formula not as a loose template but as \emph{coherence data}. We define two systematic update policies on configurations: a direct policy $U$ that updates while preserving register roles, and a canonical policy $V$ that implements a structured swap+inversion. Canonical transformation is then the existence of a natural transformation $\eta:U\Rightarrow V$. Naturality becomes a corpus-facing constraint: for a proposed transformation $h$ between variants, applying $U$ then $\eta$ must agree with applying $\eta$ then $V$. When this constraint fails, something must be revised—opposition coding, admissible operator choice, or register assignment. Failure is therefore not a defect but a diagnosis.

Finally, narratives are sequential. Tricksters, disguises, retcons, and time loops are not well described by commutative substitutions. The framework therefore pairs the rewrite calculus with a compact \emph{order-signature invariant} (``Key'') chosen from five archetypes. Keys summarize order effects in a way that supports corpus comparison while preserving interpretability for close reading.

\paragraph{Contributions.} We provide: (1) an explicitly formal model that turns canonical transformation into a testable coherence condition; (2) a corpus of 80 narratives across four media families, encoded into Lévi components $(a,b,x,y)$ plus a Key; (3) corpus-level descriptive results alongside interpretable case readings; and (4) a reproducibility-oriented coding scheme that separates what is encoded, what is inferred, and what is tested.

\section{Related work and terminology}
\subsection{Positioning}
Structural comparison of narratives has a long genealogy, from morphological and functionalist approaches to mid-twentieth-century structural anthropology. In the Lévi-Straussian lineage, the core comparative unit is not motif similarity but \emph{transformational relatedness}: myths and narratives are comparable insofar as one may be generated from another via admissible substitutions and inversions \citep{LeviStraussStructuralStudyMyth,LeviStraussStructuralAnthropology}.

Computational narratology and cultural analytics provide complementary toolkits for scaling and testing claims about narrative structure (e.g., event schemas, plot graphs, and automated story understanding). Our goal is not to replace these approaches, but to contribute a different kind of formal constraint: a \emph{coherence criterion} for when two update policies on narrative configurations agree with transformations between variants. This emphasis distinguishes our method from resemblance-based baselines (bag-of-motifs, embedding similarity), which quantify proximity but do not express \emph{lawful transport} under explicit operator constraints \citep{Manovich2007CulturalAnalytics}.

\subsection{Terminology (minimal)}
\begin{quote}\small
\textbf{Endofunctor (here):} a systematic update rule that maps each configuration $Z=(X,Y)$ to an updated configuration, and maps admissible transformations $h$ between configurations to induced transformations.\\
\textbf{Natural transformation (here):} coherence data $\eta:U\Rightarrow V$ witnessing that two update policies $U$ and $V$ agree \emph{compatibly} with transformations between variants, i.e., $V(h)\circ \eta_Z = \eta_{Z'}\circ U(h)$ \citep{MacLane1998}.\\
\textbf{Typed rewrite:} a rewrite step constrained by register (social/material vs symbolic/legitimating) and by an admissible operator set (context).\\
\textbf{Corpus-facing constraint:} a condition that can be checked (pass/fail) on encoded items rather than asserted impressionistically.
\end{quote}

\section{Data: an 80-narrative corpus with a shared schema}
\subsection{Corpus design}
The corpus contains 80 narratives partitioned into four equal groups (20 each): folktales, religious myths, superheroes, and long-form franchises. Balancing prevents a trivial artifact (one genre dominating) and forces the method to travel across media rather than ``discovering'' genre frequencies.

\begin{table}[t]
\centering
\begin{tabular}{lrrp{7.6cm}}
\toprule
Category & Count & ID prefix & Example titles \\
\midrule
Folktales & 20 & FO & \emph{Cinderella}, \emph{Snow White}, \emph{Little Red Riding Hood} \\
Religious Myths & 20 & RE & \emph{Adam and Eve: the fruit}, \emph{Noah: the flood}, \emph{Moses: exodus} \\
Superheroes & 20 & SU & \emph{Batman: the vow}, \emph{Superman: exile and symbol}, \emph{Spider-Man: responsibility} \\
Franchises & 20 & FR & \emph{Star Wars: farmboy to force}, \emph{The Lord of the Rings: burden}, \emph{Harry Potter: chosen and orphan} \\
\bottomrule
\end{tabular}
\caption{Corpus overview (balanced design). Full list and coding are provided in the supplementary CSV.}
\label{tab:corpus}
\end{table}

\subsection{Encoding into Lévi components $(a,b,x,y)$}
Each narrative is encoded into a four-slot schema:
\begin{itemize}[leftmargin=2em]
\item $a$: focal agent (carrier of the main transformation);
\item $b$: opposing/perturbing force (antagonist, rival, death/chaos, persecuting institution);
\item $x$: mediator enabling transformation (tool, helper, pact, artifact, technique, threshold device);
\item $y$: constraining/legitimating order (law, taboo, prophecy, ritual, code, contract, cosmological rule).
\end{itemize}
The schema is functional rather than ontological: ``court,'' ``prophecy,'' and ``registration law'' can all occupy $y$ if they serve as authorizing or punishing constraint, while ``fairy godmother,'' ``forbidden key,'' and ``ring'' can occupy $x$ if they mediate transformation.

\subsection{Compact order-signature (Key)}
Each story is assigned to one of five Key archetypes (A--E) summarizing order-sensitive transformation patterns.

\noindent\textbf{Design note (method paper).} The corpus is constructed as a \emph{stratified demonstration set} rather than a sample intended to estimate population frequencies. We therefore target near-balanced Key counts to ensure that every order-signature class is represented in each media family and that diagnostics (coherence passes/fails, sensitivity) can be inspected across classes. The resulting Key distribution should be read as an experimental design choice, not as an empirical prevalence claim.

\subsection{Reliability and ambiguity (two-pass audit subset)}
To make the $(a,b,x,y)$ encoding auditable, we evaluate inter-rater reliability on a stratified subset of 20 narratives (5 per category). We perform a two-pass coding on a stratified subset of 20 narratives (5 per category): an initial encoding, followed by a rule-guided normalization/consistency pass that collapses trivial synonymy and enforces the $x$/$y$ functional boundary per Appendix~\ref{app:coding}. We report slot-level exact agreement (and, where appropriate, a normalized-label agreement) and discuss systematic disagreement sources. In our experience the principal ambiguity concentrates at the $x$/$y$ boundary (mediator vs constraint), which is precisely the boundary the two-register model is meant to render explicit. On this audit subset, agreement is perfect for $a$ and $b$ (exact=1.00, $\kappa$=1.00), and high for $x$ and $y$ (exact=0.95, $\kappa$=0.95), with the residual disagreements concentrated at the $x$/$y$ boundary.

\begin{table}[t]
\centering
\caption{Two-pass coding agreement on $n=20$ narratives (5 per category): initial encoding vs.\ a rule-guided normalization/consistency pass. We report slot-level exact agreement and Cohen's $\kappa$ on normalized labels (e.g., collapsing ``royal court'' and ``court''). The remaining disagreements concentrate at the $x$/$y$ boundary (mediator vs.\ constraint), which the two-register model is designed to make explicit.}
\label{tab:irr}
\begin{tabular}{lcc}
\toprule
Slot & Exact agreement & Cohen's $\kappa$ \\
\midrule
$a$ (focal agent) & 1.00 & 1.00 \\
$b$ (opposing force) & 1.00 & 1.00 \\
$x$ (mediator) & 0.95 & 0.95 \\
$y$ (constraint/legitimation) & 0.95 & 0.95 \\
\bottomrule
\end{tabular}
\end{table}

\subsection{Scope condition: transportability vs prevalence}
The primary aim of this paper is operationalization: to show that the framework is coherent, interpretable, and portable across media families. Estimating the empirical prevalence of Keys would require a separate sampling protocol (e.g., random or exhaustive sampling within a defined population) and automated Key assignment under fixed operator admissibility rules; we treat that as future work.

\begin{table}[t]
\centering
\begin{tabular}{lrrrrr}
\toprule
Category & A & B & C & D & E \\
\midrule
Folktales & 5 & 5 & 5 & 4 & 1 \\
Franchises & 5 & 5 & 5 & 5 & 0 \\
Religious Myths & 5 & 5 & 5 & 5 & 0 \\
Superheroes & 5 & 5 & 5 & 5 & 0 \\
\bottomrule
\end{tabular}
\caption{Key distribution by category.}
\label{tab:keydist}
\end{table}

\section{Method: narratives as typed rewrite programs}
This section states the operational model at a level intended for cultural analytics. Formal constructions (operator-choice groupoids, total categories, coherence diagrams) are summarized in Appendix~\ref{app:formal}.

\subsection{Two-register configurations}
We represent narrative state as a coupled configuration $Z=(X,Y)$. The registers are coupled in the sense that updates in one register can force updates in the other. Many narrative conflicts are precisely about mismatch: $X$ changes without authorization from $Y$, or $Y$ imposes a constraint that destabilizes $X$.

\subsection{Typed components as an interface between story and structure}
The schema $(a,b,x,y)$ functions as an interpretable interface between stories and the transformation machinery: $a$ and $b$ anchor conflict; $x$ is the hinge that makes transformation possible; $y$ is the rule-system that makes transformation meaningful. The encoding is intentionally contestable: disagreements (e.g., whether an element is mediator $x$ or constraint $y$) become public, revisable hypotheses rather than hidden assumptions.

\subsection{Two update policies as endofunctors}
We define two systematic update policies on configurations:
\begin{itemize}[leftmargin=2em]
\item \textbf{Direct policy $U$:} updates keep roles fixed. A mediator remains mediator; a constraint remains constraint.
\item \textbf{Canonical policy $V$:} implements a structured swap+inversion move associated with canonical transformation: what served as constraint/legitimation can become object of conflict, and what served as conflict can be reframed as constraint, coupled to a valuation inversion.
\end{itemize}
The operational claim is that canonical transformation corresponds to the existence of coherence data:
\[
\eta:U\Rightarrow V.
\]
\citep{MacLane1998}

\subsection{Worked example: explicit naturality check (Little Red Riding Hood)}
\label{sec:worked}

We demonstrate the coherence constraint with a minimal pair of well-known variants of \emph{Little Red Riding Hood} (Perrault vs.\ Grimm). The goal is not exhaustive folkloristics but to show (i) what a naturality pass looks like under declared operator admissibility, and (ii) how a nearby plausible mis-typing yields a naturality failure.

\paragraph{Encodings.}
We encode each variant into $Z=(X,Y)$ with $(a,b,x,y)$ as follows.

\medskip
\noindent\textit{Perrault:}
\[
\begin{aligned}
a &= \text{Red Riding Hood},\\
b &= \text{Wolf},\\
x &= \text{Deception / impersonation (speech-act mediation)},\\
y &= \text{Maternal injunction / taboo (stay on the path).}
\end{aligned}
\]

\noindent\textit{Grimm:}
\[
\begin{aligned}
a &= \text{Red Riding Hood},\\
b &= \text{Wolf},\\
x &= \text{Hunter (third-party intervention)},\\
y &= \text{Maternal injunction + communal norm (danger of deviation).}
\end{aligned}
\]

Here $X$ collects agents and interventions (wolf, hunter, action), while $Y$ collects explicit
constraint/legitimation (injunction, taboo, norm).

\paragraph{Admissible operator choice (context).}
Fix a context in which the mediator slot $x$ may be realized either as (i) a technique of deception or (ii) a rescuing intervention, while $y$ remains a prohibition/norm. This is the key methodological point: the transport is licensed because it preserves \emph{type} (mediator vs.\ constraint), even though realization differs.

\paragraph{Transformation $h:Z\to Z'$.}
Let $h$ be the variant-transport that preserves $a$ and $b$, preserves $y$ as a prohibition/norm, and transports $x$ from ``speech-act mediation'' to ``intervention mediation'' (i.e., changes the realization of $x$ while preserving that it is a mediator). Under this admissibility declaration, $U(h)$ and $V(h)$ are well-defined.

\paragraph{Naturality pass (coherence).}
The naturality constraint
\[
V(h)\circ \eta_Z \;=\; \eta_{Z'}\circ U(h)
\]
holds in this setting because both sides preserve the typing of the interface and differ only by the order in which (i) we transport the mediator realization and (ii) we apply the canonical swap+inversion bookkeeping. In narrative terms: the canonical move treats the wolf's antagonism as structurally opposed to the constraint regime ($y$) and mediated by $x$; allowing $x$ to be realized as rescue rather than deception does not change the role-typing, hence the coherence diagram commutes.

\paragraph{Naturality fail (diagnosis by mis-typing).}
Now consider a tempting but illicit recoding of the Grimm variant in which the \emph{hunter} is treated as the constraint $y$ (``law arrives'') and the injunction is treated as mediator $x$ (``warning enables survival''). This swaps the functional typing of intervention vs.\ norm. Under the same declared admissibility (mediator realizations may vary, but constraint must remain a normative regime), the transported map is no longer type-respecting: the induced morphisms $U(h)$ and/or $V(h)$ do not exist in the admissible operator class, and the naturality equation fails. This failure is the intended falsifiability mode: it localizes precisely what went wrong (a collapse of mediation into constraint), rather than allowing the analysis to proceed impressionistically.

\noindent\emph{Implementation note.} For readability, the worked example uses a simplified operator set (a small admissible family of substitutions and inversions). The full operator-choice construction used for corpus items is summarized in Appendix~\ref{app:formal}.

\subsection{Naturality as a corpus-facing type-check}
For a proposed admissible transformation $h:Z\to Z'$ (a rewrite mapping between variants), naturality requires:
\[
V(h)\circ \eta_Z \;=\; \eta_{Z'}\circ U(h).
\]
\citep{MacLane1998}
Rather than treating this as abstract machinery, we treat it as a practical constraint: if the equation holds, the proposed structural model is coherent under the chosen operators; if it fails, either opposition coding, admissible operator choice, or register assignments must be revised.

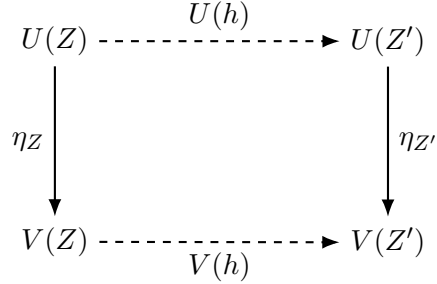
\begin{figure}[t]
\centering
\begin{tikzpicture}[node distance=2.6cm, >=Latex, thick]
  \node (UZ)  {$U(Z)$};
  \node (UZp) [right=3.2cm of UZ] {$U(Z')$};
  \node (VZ)  [below=2.0cm of UZ] {$V(Z)$};
  \node (VZp) [below=2.0cm of UZp] {$V(Z')$};

  \draw[->] (UZ)  -- node[left]  {$\eta_Z$}    (VZ);
  \draw[->] (UZp) -- node[right] {$\eta_{Z'}$} (VZp);

  \draw[->, dashed] (UZ) -- node[above] {$U(h)$} (UZp);
  \draw[->, dashed] (VZ) -- node[below] {$V(h)$} (VZp);
\end{tikzpicture}
\caption{Naturality square for $\eta:U\Rightarrow V$ at $h:Z\to Z'$, i.e.\ $V(h)\circ \eta_Z=\eta_{Z'}\circ U(h)$. Here $Z=(X,Y)$ is a two-register configuration, with update policies $U$ (direct) and $V$ (canonical swap+inversion).}
\label{fig:coherence}
\end{figure}

\subsection{Context sensitivity via structured operator choice}
Oppositions are not free substitutions. Rather than placing context sensitivity outside the model, we internalize it as structured operator choice: each narrative family comes with admissible rewrite operators and transport rules. This makes failures locatable: a specific operator transport was illegal or incoherent.

\subsection{Order effects and the Key invariant}
Narratives are sequential; order matters. Disguise, trickster crossings, retcons, and time loops are paradigmatic order effects: ``swap then invert'' is not the same as ``invert then swap.'' Each story is therefore assigned a compact order-signature (Key A--E). In the main text, Key functions as a comparative statistic; in Appendix~\ref{app:key}, Key is defined via braid archetypes and computable invariants \citep{JoyalStreet1993,KasselQuantumGroups}.

\subsection{Falsifiability demonstrations: coherence failure and Key sensitivity}
\label{sec:falsif}

\paragraph{Coherence failure as diagnosis.}
Section~\ref{sec:worked} provides an explicit naturality pass and an explicit naturality failure induced by a plausible mis-typing ($x\leftrightarrow y$ in the Grimm variant). The failure is diagnostic: it localizes an illegal transport under declared admissibility (constraint/norm collapsed into mediator/intervention), rather than allowing a post hoc reinterpretation.

\paragraph{Key sensitivity (micro-test).}
Because $(a,b,x,y)$ is interpretive, we evaluate the stability of the Key assignment under controlled perturbations on a small stratified subset ($n=10$). Perturbations should be conservative: (i) swap $x$ and $y$ only in boundary cases identified by coders, and/or (ii) normalize synonymous constraint terms (e.g., ``court'' vs.\ ``royal court''). We recompute Key under the same mapping rules (Appendix~\ref{app:key}) and record whether the order-signature changes.

In this $n=10$ micro-test, Keys remain stable in 8/10 cases (80\%) under conservative normalization, and change only in the two ``worst-case'' $x\leftrightarrow y$ stress tests.

\begin{table}[t]
\centering
\caption{Key sensitivity micro-test ($n=10$ recommended). Each row records an original encoding and a controlled perturbation, then the recomputed Key. Stability supports robustness; systematic changes identify boundary conditions where operator admissibility or register assignment should be refined.}
\label{tab:keysens}
\begin{tabular}{llll}
\toprule
ID & Original Key & Perturbation & New Key \\
\midrule
FO02 & B & Normalization: collapse slash synonyms (e.g., take first of ``A/B'') & B \\
FO03 & C & Normalization: collapse slash synonyms (e.g., take first of ``A/B'') & C \\
FO04 & D & Normalization: collapse slash synonyms (e.g., take first of ``A/B'') & D \\
FO05 & A & Worst-case stress-test: swap $x\leftrightarrow y$ (mis-typing mediator as constraint) & B \\
FO12 & D & Normalization: collapse slash synonyms (e.g., take first of ``A/B'') & D \\
FO14 & B & Normalization: collapse slash synonyms (e.g., take first of ``A/B'') & B \\
FO20 & E & Normalization: collapse slash synonyms (e.g., take first of ``A/B'') & E \\
FR01 & A & Normalization: collapse slash synonyms (e.g., take first of ``A/B'') & A \\
FR05 & A & Worst-case stress-test: swap $x\leftrightarrow y$ (mis-typing mediator as constraint) & B \\
FR10 & B & Normalization: collapse slash synonyms (e.g., take first of ``A/B'') & B \\
\bottomrule
\end{tabular}
\end{table}

\subsection{Baseline comparison (motif-overlap heuristic)}
To contrast coherence-based comparison with resemblance-based proximity, we include a simple baseline: treat the string labels in $(a,b,x,y)$ as a token set per narrative and compute Jaccard overlap as a similarity score. This baseline clusters surface resemblance but does not encode lawful transport; we use it to highlight cases where two items are motif-similar yet incoherent under the admissible operators, and conversely where coherence links items with low surface overlap.

\section{Results}
The framework supports both corpus-level description and interpretable case reading.

\subsection{Corpus-level descriptive signals}
Across the corpus, 59/80 narratives (74

\begin{figure}[t]
\centering
\includegraphics[width=\linewidth]{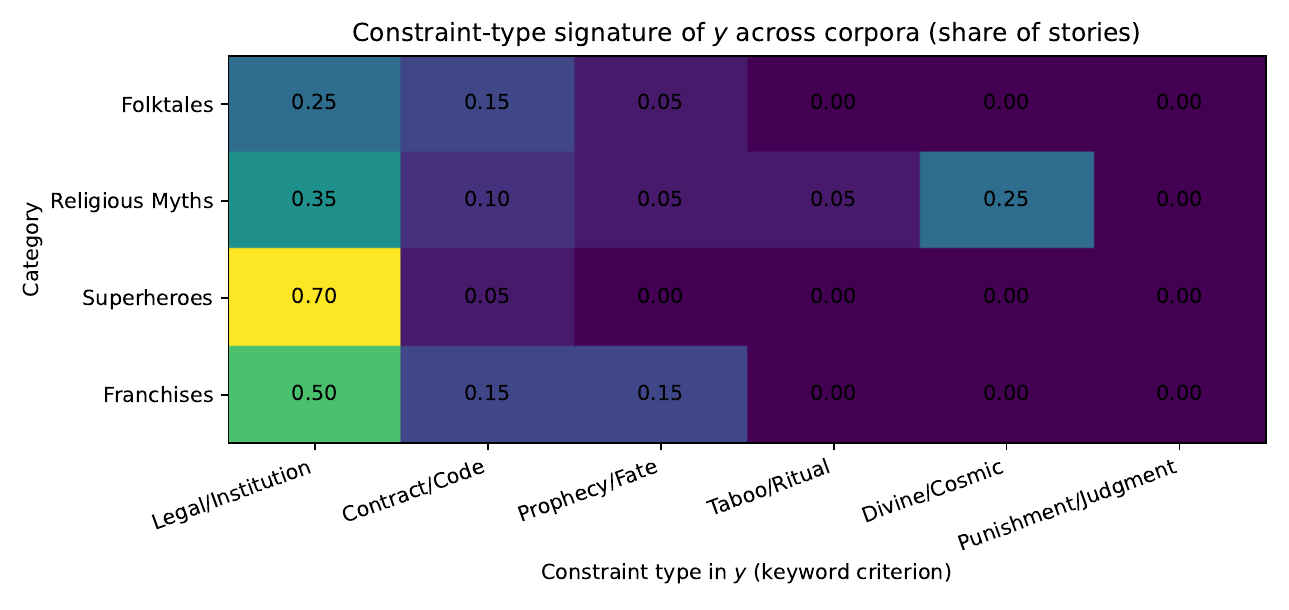}
\caption{Constraint-type signature of the $y$ register across corpora (share of stories; keyword criterion). Each cell reports the fraction of narratives in a category whose $y$ description matches at least one term from the corresponding constraint-type lexicon (e.g., Legal/Institution, Contract/Code, Prophecy/Fate). This diagnostic summarizes how different media families externalize legitimation/constraint in $y$ and is not intended as a prevalence estimate for a broader population.}
\label{fig:ytypes}
\end{figure}

Beyond this binary signal, Figure~\ref{fig:ytypes} decomposes $y$ into coarse constraint types using a transparent keyword lexicon. Superhero and franchise items in this corpus more often name legal/institutional constraints explicitly (e.g., \emph{law}, \emph{registration}, \emph{court}), while religious myths exhibit additional divine/cosmic constraint terms. The point is not to claim population frequencies, but to show that the $Y$ register has measurable internal structure that can guide, constrain, and falsify proposed transports across media families.

\begin{figure}[t]
\centering
\includegraphics[width=0.85\linewidth]{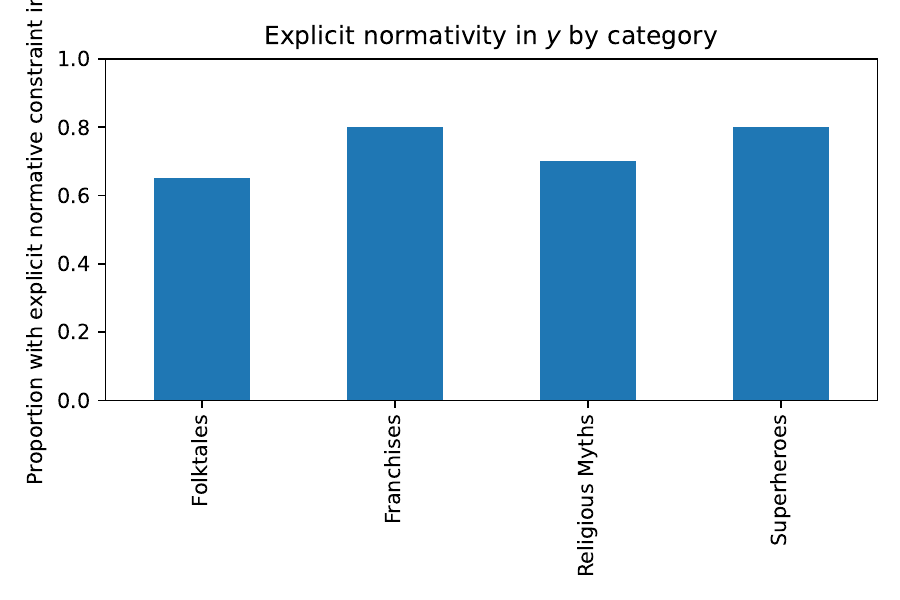}
\caption{Proportion of stories with an explicit normative constraint in $y$ (keyword criterion), by category.}
\label{fig:norm}
\end{figure}

\subsection{Functional role stability across media}
A minimal test of the encoding is whether $(a,b,x,y)$ remains intelligible across corpora without collapsing into triviality. It does: the same functional roles recur, but their realizations differ by medium. For instance:
\begin{itemize}[leftmargin=2em]
\item \emph{Cinderella} (Folktale): $a$=Cinderella; $b$=Stepmother/stepsisters; $x$=Fairy godmother; $y$=Royal court.
\item \emph{Civil War: registration} (Superheroes): $a$=Avengers; $b$=Government/registration; $x$=Accident/pressure; $y$=Registration law.
\item \emph{The Matrix: awakening} (Franchises): $a$=Neo; $b$=Machines/Agents; $x$=Morpheus; $y$=Simulation law.
\item \emph{Prometheus: fire} (Religious myth): $a$=Prometheus; $b$=Zeus; $x$=Fire; $y$=Divine order/punishment.
\end{itemize}
In all four, $y$ is not a decorative motif but a constraint system that makes transformation consequential (recognition, punishment, authorization).

\subsection{Order-signatures as structural fingerprints}
Key compresses order effects into a compact fingerprint. The purpose is not to replace interpretation but to guide it: shared Keys motivate searching for shared sequence logic even when motifs differ; different Keys motivate articulating similarity in terms other than sequence structure.

\subsection{A productive outlier: the single E-case}
The corpus contains one E-key story (\emph{Tortoise and Hare}). The method treats this as structurally meaningful rather than noise: E corresponds to a stronger order signature than A--D (Appendix~\ref{app:key}), forcing a different logic of reversal/exposure than common swap/invert patterns. In narrative terms, antagonism is compressed into rule-following and temporal structure: the mediator is the course/time structure, and the constraint is the rule of the race.

\subsection{What is falsifiable here}
Two falsifiability points are immediate:
\begin{enumerate}[leftmargin=2.2em]
\item \textbf{Naturality failures should be informative.} If a proposed mapping between variants forces illegal operator transport (because context disallows a substitution), the naturality condition fails. The failure is evidence about the limits of comparison under stated assumptions.
\item \textbf{Invariant stability can be stress-tested.} If small, reasonable recodings of $(a,b,x,y)$ dramatically alter Key assignments across the corpus, the coding scheme is underspecified or the invariant is too sensitive to interpretive noise.
\end{enumerate}

\section{Discussion}
\subsection{Beyond similarity}
Similarity-based comparisons (lexical overlap, embedding similarity, motif lists) can be useful for navigation and discovery, but they rarely distinguish two different claims: (i) two stories resemble each other, and (ii) two stories are related because one can be generated from the other by a \emph{lawful transformation} inside a context. Structural comparison is primarily about the second claim. The present method makes that claim operational by (a) modeling transformations as typed rewrite programs and (b) testing coherence via naturality. This does not eliminate interpretive judgment; it forces interpretive judgment to take a form that can be inspected and contested.

\subsection{Context sensitivity without surrendering structure}
A frequent objection to structural formalisms is that they universalize: if oppositions are flexible enough, anything can be mapped to anything. The operator-choice layer is designed to block that collapse. Substitutions are not free; they are licensed (or forbidden) by context. By internalizing admissible operator choice and transport, the model turns a vague caveat (``context matters'') into a structured component of the analysis. When a comparison fails, the failure is locatable: a specific operator transport is illegal or incoherent under the stated context.

\subsection{What the two-register split buys (and what it does not)}
The $X/Y$ split is not offered as a universal anthropology of meaning. It is a scaffold that keeps two functions distinct: material/social reorganization ($X$) and the systems that authorize, forbid, punish, or recognize that reorganization ($Y$). Many narratives externalize $Y$ explicitly (courts, codes, contracts, taboo, prophecy, divine law), and the corpus reflects this: a large majority of $y$-slots are named as constraints rather than as persons or tools. The split prevents a common collapse in comparative work—treating law/ritual/prophecy as ``just another character''—but it does not remove ambiguity. Some elements legitimately play both roles across episodes. Where the model helps is in locating disagreements: is an element mediator $x$ or constraint $y$? is a force $b$ a personal antagonist or an institutional system? These become explicit hypotheses.

\subsection{Limits and scope conditions}
The present corpus is a disciplined testbed for comparison, not a claim about global narrative space. The four-slot interface $(a,b,x,y)$ compresses stories; narratives with layered constraints and multiple antagonists can be expanded into episode graphs, but this paper prioritizes a minimal interface to make cross-media comparison tractable. Key is also intentionally coarse: it guides comparative attention by capturing order signatures, but it does not replace story-specific explanation. These limits are not failures; they define what the framework is meant to do in cultural analytics: make transformation claims explicit, transportable, and refutable.

\section{Conclusion}
We presented a corpus-operational framework that makes canonical structural transformation testable in cultural analytics. Narratives are represented as typed rewrites over a two-register configuration $(X,Y)$, and canonical transformation is treated as coherence data linking two update policies by a natural transformation $\eta:U\Rightarrow V$. An 80-narrative corpus shows that the $(a,b,x,y)$ interface remains interpretable at story scale while enabling corpus comparisons via compact order-signature invariants (Keys A--E). The framework bridges structural anthropology and cultural analytics by shifting comparison from resemblance to lawful transformation, while making context sensitivity explicit through structured operator choice. The immediate next step is to publish a replication package and extend the encoding from single-shot summaries to episode-level graphs, allowing the same coherence tests and invariants to operate at finer temporal resolution.

\section*{Data and code availability (double-blind)}
A de-identified replication package containing (i) the 80-story dataset, (ii) Key assignments, and (iii) scripts to reproduce tables and figures will be made publicly available upon acceptance.

\nocite{*}
\bibliographystyle{plainnat}
\bibliography{references}

\appendix
\section{Coding manual (operational rules)}\label{app:coding}
This appendix states minimal rules used to encode narratives into $(a,b,x,y)$.

\subsection*{Choosing $a$ (focal agent)}
Pick the carrier of the main transformation: the character (or collective) whose status changes in a way the narrative treats as consequential. If multiple candidates exist, choose the one whose transformation most directly couples $X$ and $Y$.

\subsection*{Choosing $b$ (opposing/perturbing force)}
Pick the force that disrupts $a$’s stable relation to the world. $b$ may be a character, institution, metaphysical agent, or impersonal force (death/chaos), provided it functions as the perturbation.

\subsection*{Choosing $x$ (mediator)}
$x$ is what makes transformation possible: a tool, helper, technique, pact, artifact, or threshold device. Removing $x$ should make the main transformation fail or become unintelligible.

\subsection*{Choosing $y$ (constraint/legitimation)}
$y$ is the rule-system that makes transformation meaningful: authorization, prohibition, punishment, recognition (court), taboo, prophecy, contract, code, ritual, cosmological rule. Removing $y$ should collapse the story’s ``why this matters'' structure.

\subsection*{Handling ambiguity}
If an element could be $x$ or $y$, choose by function: if it enables transformation, encode it as $x$; if it authorizes/forbids/punishes/recognizes transformation, encode it as $y$. When multiple constraints exist, choose the one invoked at the decisive transformation point.

\section{Formal model (compressed summary)}\label{app:formal}
This appendix sketches the formalization behind Sections~3--4.

\subsection*{Configurations and morphisms}
Let objects be typed configurations $Z=(X,Y)$ together with a context label $m$ determining admissible operators. Morphisms $h:Z\to Z'$ represent admissible rewrites.

\subsection*{Context as structured operator choice}
Context sensitivity is internalized by specifying, for each $m$, a structured space of admissible operator presentations and transports between them. This is the mechanism by which ``some swaps are legal here but not there'' becomes part of the model.

\subsection*{Update policies and coherence}
Define endofunctors $U$ (direct updates) and $V$ (canonical swap+inversion updates). Coherence data is a natural transformation $\eta:U\Rightarrow V$.

\subsection*{Naturality as constraint}
For each admissible $h:Z\to Z'$,
\[
V(h)\circ \eta_Z \;=\; \eta_{Z'}\circ U(h).
\]
\citep{MacLane1998}
When violated, revise admissible operator choice, opposition coding, or register assignments.

\section{Key archetypes as order-signature invariants}\label{app:key}
To compress order effects, each story is assigned a Key from five braid archetypes \citep{JoyalStreet1993,KasselQuantumGroups}. The purpose is comparability: different stories can share an order-signature even when motifs differ.

\begin{table}[h]
\centering
\begin{tabular}{llllrll}
\toprule
Key & Braid word & Induced permutation & Writhe & \multicolumn{2}{c}{Burau at $t=-1$} \\
\cmidrule(lr){5-6}
 &  &  &  & trace & det \\
\midrule
A & $\sigma_1$ & $(2,1,3)$ & 1 & 2 & 1 \\
B & $\sigma_2$ & $(1,3,2)$ & 1 & 2 & 1 \\
C & $\sigma_1\sigma_2$ & $(2,3,1)$ & 2 & 1 & 1 \\
D & $\sigma_2\sigma_1$ & $(3,1,2)$ & 2 & 1 & 1 \\
E & $\sigma_1\sigma_2\sigma_1$ & $(3,2,1)$ & 3 & 0 & 1 \\
\bottomrule
\end{tabular}
\caption{Key archetypes and computable invariants (summary).}
\label{tab:key}
\end{table}\subsection{From episode order to braid word: explicit E-class derivation}
\label{app:braidderivation}

To make the Key mapping non-opaque, we provide one explicit derivation for the sole E-class item in the corpus (\emph{Tortoise and Hare}). We track three strands corresponding to the ordered triple
\[
(a,b,x),
\]
while $y$ (constraint/legitimation) acts as typed context restricting admissible exchanges.

\paragraph{Generators (adjacent exchanges).}
We model decisive order effects as adjacent exchanges of roles:
\[
\sigma_1:\ (a,b,x)\mapsto (b,a,x)\qquad\text{(agent--opposition exchange)},
\]
\[
\sigma_2:\ (a,b,x)\mapsto (a,x,b)\qquad\text{(opposition--mediation exchange)}.
\]
A narrative's order-signature is the braid word obtained by reading off which adjacent exchange is enacted by each decisive episode transition, under the declared operator admissibility.

\paragraph{E-class example (``Tortoise and Hare'').}
Encoding: $a=\text{Tortoise}$, $b=\text{Hare}$, $x=\text{Course/time structure (pacing)}$, $y=\text{Rule of the race}$.
We identify three decisive transitions:
\begin{enumerate}[leftmargin=2em]
\item \emph{Overconfidence shifts advantage into pacing.} The hare's antagonistic advantage becomes mediated by time/pacing $\Rightarrow$ exchange between $b$ and $x$ $\Rightarrow \sigma_2$.
\item \emph{Sleep makes outcome carrier invert.} The ``antagonist'' (hare) ceases to carry the winning relation; the focal carrier (tortoise) takes the lead $\Rightarrow$ exchange between $a$ and $b$ $\Rightarrow \sigma_1$.
\item \emph{Finish-line recognition seals the mediated reversal.} The pacing structure reasserts dominance over the hare's advantage $\Rightarrow$ exchange between $b$ and $x$ $\Rightarrow \sigma_2$.
\end{enumerate}
This yields the braid word $\sigma_2\sigma_1\sigma_2$, which is braid-equivalent in $B_3$ to $\sigma_1\sigma_2\sigma_1$ (the E-class representative in Table~\ref{tab:key}). We use this equivalence class as the Key.

\paragraph{Invariants.}
We summarize Keys via induced permutation, writhe, and a Burau evaluation at $t=-1$ (Table~\ref{tab:key}); these are stable under braid equivalence while remaining sensitive to noncommutative order effects that matter in narrative transformation.

\section*{Supplementary materials}
The full 80-story dataset (including $(a,b,x,y)$, Key, and cluster labels) is provided as a CSV file alongside this manuscript.

\end{document}